\documentclass[
reprint,%
aps,%
prl,%
groupedaddress,%
superscriptaddress,%
twocolumn,%
assume,%
amsmath,%
floatfix
]{revtex4-2}

\usepackage{graphicx}
\usepackage[
  colorlinks=true,
  urlcolor=blue,
  linkcolor=blue,
  citecolor=blue
]{hyperref}
\usepackage{siunitx,braket,amssymb,gensymb,ulem,color}

\newcommand{\rrm}[1]{\textrm{#1}}

\newcommand{\dd}[2]{\frac{\partial#1}{\partial#2}}
\newcommand{\lr}[1]{\left(#1\right)}

\usepackage{graphicx}
\usepackage{dcolumn}
\usepackage{bm}
\usepackage{ORCIDinREVTeX}

\begin{document}

\preprint{APS/123-QED}

\title{Optical skyrmions of vortex darkness}

\author{Nilo Mata-Cervera}
\orcid{0000-0001-8464-5102}
\email{nilo001@e.ntu.edu.sg}
\affiliation{Centre for Disruptive Photonic Technologies, School of Physical and Mathematical Sciences, Nanyang Technological University, Singapore 637371, Republic of Singapore}
\author{Deepak K. Sharma}
\orcid{0000-0002-5733-3952}
\affiliation{Institute of Materials Research and Engineering (IMRE), Agency for Science, Technology and Research (A*STAR), Singapore 138634, Republic of Singapore}
\author{Yijie Shen}
\orcid{0000-0002-6700-9902}
\affiliation{Centre for Disruptive Photonic Technologies, School of Physical and Mathematical Sciences, Nanyang Technological University, Singapore 637371, Republic of Singapore}
\affiliation{School of Electrical and Electronic Engineering, Nanyang Technological University, Singapore 639798, Republic of Singapore}
\author{Ramon Paniagua-Dominguez}
\orcid{0000-0001-7836-681X}
\affiliation{Institute of Materials Research and Engineering (IMRE), Agency for Science, Technology and Research (A*STAR), Singapore 138634, Republic of Singapore}
\author{Miguel A. Porras}
\orcid{0000-0001-8058-9377}
\affiliation{Complex Systems Group, ETSIME, Universidad Politécnica de Madrid, Ríos Rosas 21, 28003 Madrid, Spain}

\begin{abstract}
We disclose the existence of a type of optical skyrmion, Gauss-Stokes (GS) skyrmions, that is naturally present in an optical vortex around its phase singularity. Contrary to previous research with optical skyrmions, we neither shape vector beams nor superpose different spatial modes and polarizations. In GS skyrmions, the phase singularity in the transversal field of a single monochromatic beam of uniform polarization (a scalar beam) is concealed by the axial field dictated by Gauss's divergence law, giving rise to a polarization singularity of undefined polarization plane. This singularity is enclosed by a rich skyrmionic polarization texture fulfilling a topological map and covering all the states of transverse-axial polarization. In our experiment, we facilitate the observation of a GS skyrmion with the predicted features using focused fields with enhanced axial component. 
\end{abstract}

\maketitle

Optical vortices (OVs) are among the most fascinating phenomena in wave optics, sparking significant research advancements in recent years~\cite{coullet1989optical,shen2019optical,forbes2024orbital}. They are characterized by an azimuthal variation of the phase, resulting in a corkscrew-shaped wavefront as they propagate. The hallmark of OVs is the \textit{phase singularity}, a point hidden in the darkness where the phase is undefined and the intensity drops to zero~\cite{berry2023singularities}.
As we trace a closed loop around the vortex singularity the phase wraps a number of full $[0,2\pi]$ cycles specified by the topological charge (TC). This structure appears as a fundamental solution of the scalar wave equation for monochromatic light, and it may be embedded in light beams of different shapes such as Bessel or Laguerre-Gauss (LG) beams~\cite{forbes2021structured,rubinsztein2016roadmap} carrying orbital angular momentum (OAM)~\cite{Allen1992PRA,indebetouw1993optical}. OVs have become the cornerstone of diverse branches of optics such as optical communications~\cite{yan2014high,zhao2015capacity,wang2022orbital}, optical manipulation~\cite{Cheng2025,otte2020optical,simpson1996optical,berry2013physical,yang2021optical}, super-resolution microscopy~\cite{vicidomini2018sted,wang2017improved}, to name a few. 

Rich phenomena arise when various OVs are combined with different polarization features. For instance, when two circularly polarized vortices are superposed with opposite handedness and TCs, the electromagnetic fields feature azimuthal or radial 
polarization distributions, giving rise to a \textit{polarization singularity}~\cite{mclaren2015measuring,liu2016generation}. They constitute an example of the so-called vector beams, where all linear polarization states at the equator of the Poincaré sphere (PS) are present in a transverse plane. A more sophisticated vector texture is the skyrmion, a vector field spanning the entire surface of a parametric sphere and fulfilling a topological map. Although their origin goes back to magnetic materials~\cite{magnetic_skyrmions_1,magnetic_skyrmions_2,magnetic_skyrmions_3}, skyrmions are now realized with optical vectors such as the electromagnetic fields~\cite{zeng2024tightly,liu2022disorder}, the Poynting vector~\cite{wang2024topological}, the spin angular momentum~\cite{spin_sk,spin_sk_2}, to cite a few. Skyrmions carry an integer, topologically invariant scalar quantity known as skyrmion number~\cite{OpticalSk_NP,theory_skyrmions_1,theory_skyrmions_2,shen2021topological} measuring the number of times the vector field wraps the sphere. Owing to the flexibility of current light shaping technologies~\cite{shen2022generation,hakobyan2024unitary,he2024optical,kerridge2024optical,lin2024chip}, the most common way to construct them is with the Stokes vector: two different spatial modes 
are superimposed in orthogonal polarizations such that the transverse plane maps all the surface of the PS~\cite{theory_skyrmions_1,theory_skyrmions_2,shen2021topological}. 

\begin{figure*}[t!]
    \centering
    \includegraphics[width=\linewidth]{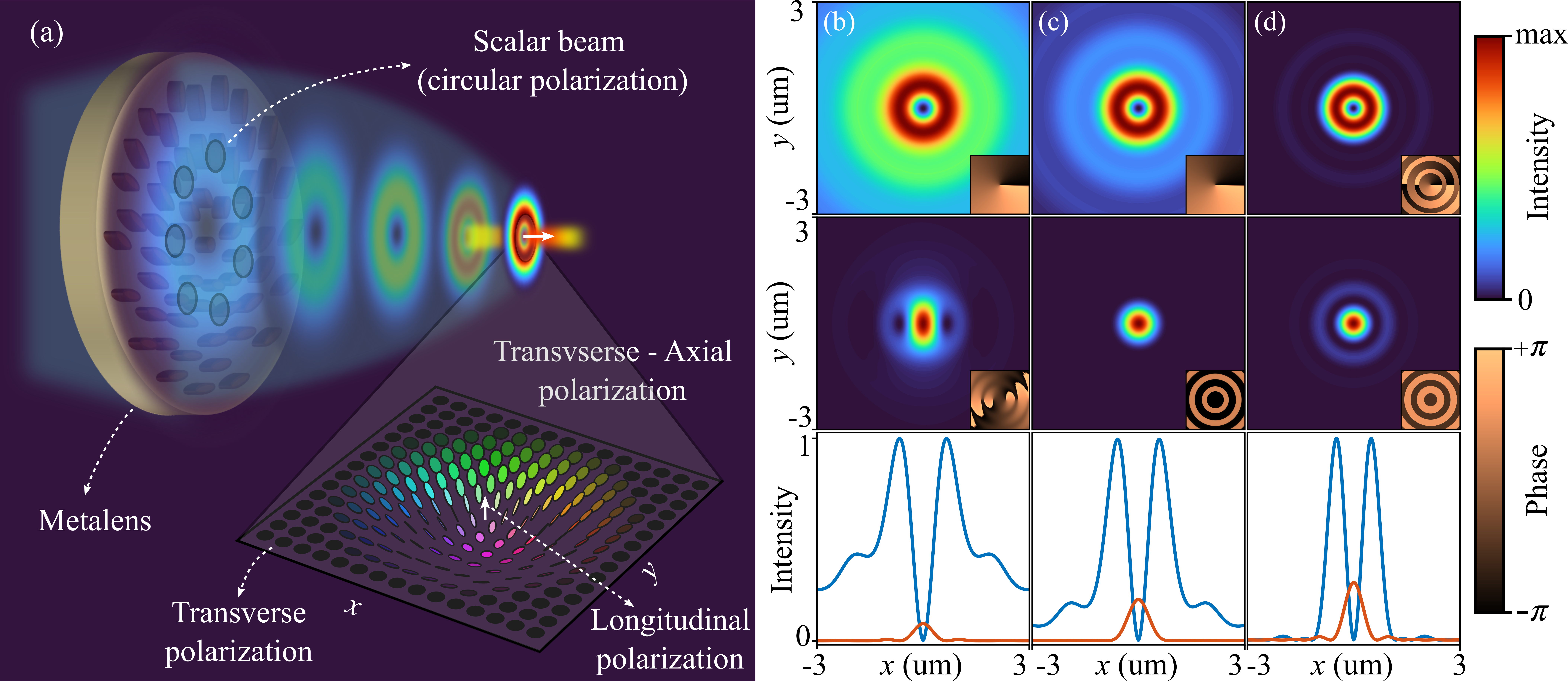}
    \caption{(a) GS skyrmions emerging from the vortex singularity: a scalar beam develops an observable skyrmion structure of TA polarization when focused. Focused profiles of input linearly-polarized (b) and RCP (c) EVBs (\ref{eq:exploding_profile}) with $s=1$,  $\mu=0.2$, $\sigma=\SI{25}{\micro\metre}$ and (d) input RCP-polarized uniform vortex with $s=1$. Top row shows their transverse intensity and phase (inset), same for the axial field in the middle row. The comparison between axial (red) and transverse (blue) intensities at $y=0$ is depicted in the lower row, the intensity scale is normalized to the peak transverse intensity. In all plots $f=\SI{1500}{\micro\metre}$, $\rrm{NA}=0.5$, $\lambda=\SI{735}{\nano\metre}$, and the results are calculated using Rayleigh-Sommerfeld vector diffraction integrals \cite{sommerfeld}.}
    \label{fig1}
\end{figure*}

In this Letter we demonstrate that skyrmion is a fundamental feature of OVs, 
an attribute that emerges naturally from their transverse-axial (TA) polarization features~\cite{afanasev2023nondiffractive}. In other words, one scalar OV is a skyrmion itself. It differs substantially from standard Stokes skyrmions in that they are not constructed \textit{ad libitum}, instead they are inherent to OVs due to Maxwell's equations. As they arise in an OV from Gauss's divergence, we refer to them as Gauss-Stokes (GS) skyrmions. The darkness of the phase singularity is cloaked by the longitudinal field yielding a \textit{singularity of the plane of polarization}, or $z-$point following Nye's nomenclature~\cite{nye1983lines,nye1983polarization}. For transverse circular polarization (CP), this point is surrounded by a rich TA polarization texture, in which the polarization plane rotates azimuthally when we move \textit{around the singularity}, and inclines from the transverse plane towards the optical axis when we go \textit{towards the singularity}. The resulting polarization texture fulfills a skyrmion map from a 2-sphere to the transverse plane $(S^2\rightarrow\mathbb{R}^2)$ and carries an integer skyrmion number [Fig. \ref{fig1}(a)]. 
Of course, the skyrmions in this paper neither belong to 3-sphere nor to 4-sphere stereographic projections~\cite{sugic2021particle,marco2022optical} since the transverse polarization is strictly uniform. We first discuss the structure of GS skyrmions and suitable conditions for their observation [Fig.\ref{fig1}(a)]. We then confirm their existence experimentally, and conclude by demonstrating their universality in OVs.

To make a GS skyrmion visible, we might resort to strong, nonparaxial focusing for a longitudinal field to emerge sharply, but enhanced longitudinal fields are also present in some paraxial beams. Paraxial focusing greatly simplifies the analysis. The focused field at the focal plane from a monochromatic scalar field $\psi_0(r)e^{is\varphi}$ with a vortex of TC $s$ and cylindrically symmetric amplitude is described by Fresnel diffraction integral, which takes the specific form \cite{PorrasEXPLODING}
\begin{eqnarray}\label{eq:FRESNELR}
 \psi(r)e^{is\varphi}=\frac{e^{\frac{ikr^2}{2f}}k}{i^{|s|+1}f}\int_0^R dr' r' \psi_0(r') J_{|s|}\left(\frac{k}{f}r r'\right) e^{is\varphi} ,
\end{eqnarray}
where $J_{|s|}(\cdot)$ is the Bessel function of first kind and order $|s|$, $k$ is the propagation constant, $f$ is the focal length and $R$ the aperture radius, satisfying $ {\rm NA} =\sin[\tan^{-1}(R/f)]\leq 0.5$ \cite{siegman1986lasers}. Two states of polarization in the transversal plane are of interest to us: linear polarization ${\boldsymbol \psi} =\psi {\bf u}_{x,y}$ along $x$ or $y$, where ${\bf u}_{x,y}$ are unit vectors along the $x$ and $y$ transverse directions, and CP ${\boldsymbol \psi} =\psi {\bf u}_{l,r}$, where ${\bf u}_{l,r}= ({\bf u}_x\pm i {\bf u}_y)/\sqrt{2}$ and the plus and minus signs stand for left- (LCP) and right- (RCP) handed CPs. These transversal components are accompanied by a typically negligible longitudinal field described in the paraxial approximation by $\psi_z = (i/k)\nabla_\perp \cdot {\boldsymbol \psi}$ \cite{Lax_method}, where $\nabla_\perp \cdot$ is the transversal divergence. The above axial component follows directly from Gauss divergence law $\nabla \cdot {\bf E}=0$ for an electric field of the form ${\bf E}={\boldsymbol \psi}e^{ikz}$ when $\partial \psi_z/\partial z$ is neglected compared to $i k\psi_z$ for a paraxial field. For linear polarization along $x$, the axial component yields
\begin{equation}
\psi_z= \frac{i}{k}\left[\frac{\partial\psi(r)}{\partial r} \cos\varphi - i s \psi(r) \frac{\sin\varphi}{r}\right] e^{is\varphi},
\end{equation}
and for CP yields
\begin{equation}\label{eq:CP_profile}
\psi_z = \frac{i}{\sqrt{2}k}\left[\frac{\partial\psi(r)}{\partial r} \mp \frac{s}{r}\psi(r)\right]
e^{i(s\pm 1)\varphi},
\end{equation}
where upper and lower signs pertain to LCP and RCP. 

A first example of non-negligible, enhanced axial component is the so-called exploding vortex beam (EVB) \cite{PorrasEXPLODING,Nilo_ACSPh}
\begin{equation}\label{eq:exploding_profile}
   \psi_0(r)e^{is\varphi}=A\frac{(r/\sigma)^{|s|}}{\left[1+(r/\sigma)^2\right]^{\mu+1}}e^{is\varphi},
\end{equation}
where $A$ is a constant and the parameter $\sigma$ scales the EVB to the desired dimensions. If $(|s|-1)/2 < \mu < |s|/2$, the EVB carries finite power in the whole transversal plane. When focused with infinite aperture ($R\rightarrow \infty$), the EVB produces an infinitely narrow and intense ring surrounded by a punctual vortex, where the axial component has infinite intensity. With finite aperture the intensity is always finite, but the ideal exploding behavior remains in an enhanced axial component compared to standard LG beams. For a particular choice of the parameters yielding ${\rm NA}=0.5$ and with $s=1$, the intensity and phase of the transverse and axial components at the focal plane are shown in Fig. \ref{fig1}(b) for linear polarization and in Fig. \ref{fig1}(c) for CP.

Another enhanced axial component appears when focusing a uniform profile with a punctual vortex at the center, described by $\psi_0(r)e^{is\varphi}$ with $\psi_0(r)=A$ if $r\le R$, zero otherwise, and $A$ is a constant.
%
%
The focused field can be evaluated from Eq. (\ref{eq:FRESNELR}) and integral 6.561.13 in \cite{integrals} or numerically,
%
%
and the same for its derivative in order to compute the axial field. 
The intensity and phase of transverse and axial components 
at the focal plane are shown in Fig. \ref{fig1}(d) for $\rrm{NA}=0.5$, $s=1$, and only for CP for brevity. The detailed structure of the 
profiles shown in Figs. \ref{fig1}(b,c,d) is discussed later on. We just note here the increasingly enlarged axial component from linear to CP, and from EVB to uniform illumination.


The skyrmions discussed here belong to the class of Stokes skyrmions. They are typically constructed by superposing two different OAM modes with orthogonal polarizations, say 
$\boldsymbol{\psi}=\psi_A{\bf u}_{A}+\psi_B{\bf u}_{B}$,
such as the resulting beam maps all the points of the PS into the transversal plane. 
At their simplest, they are shaped with $\psi_A=LG_{0,0}$, $\psi_B=LG_{l,0}$, with $\bm{u}_{A}=\bm{u}_{l}$, $\bm{u}_{B}=\bm{u}_{r}$, and $LG_{l,p}$ the LG beam of azimuthal order $l$ and radial number $p$. The skyrmion number of the normalized Stokes field $\bm{s}=(S_1,S_2,S_3)/S_0$, calculated as
\begin{eqnarray}\label{eq:Nsk_xy}
N_{\rm SK} = \!\int\! n_{\rm SK} r\rrm{d}r\rrm{d}\varphi
= \frac{1}{4\pi}\!\int\!\bm{s}\cdot\lr{\dd{\bm{s}}{r}\!\times\!\dd{\bm{s}}{\varphi}}r\rrm{d}r\rrm{d}\varphi,
\end{eqnarray}
yields $N_{\rm SK}=l$ when integrating (\ref{eq:Nsk_xy}) across the entire transversal plane \cite{theory_skyrmions_1,theory_skyrmions_2}. 

\begin{figure*}[t!]
    \centering
\includegraphics[width=\linewidth]{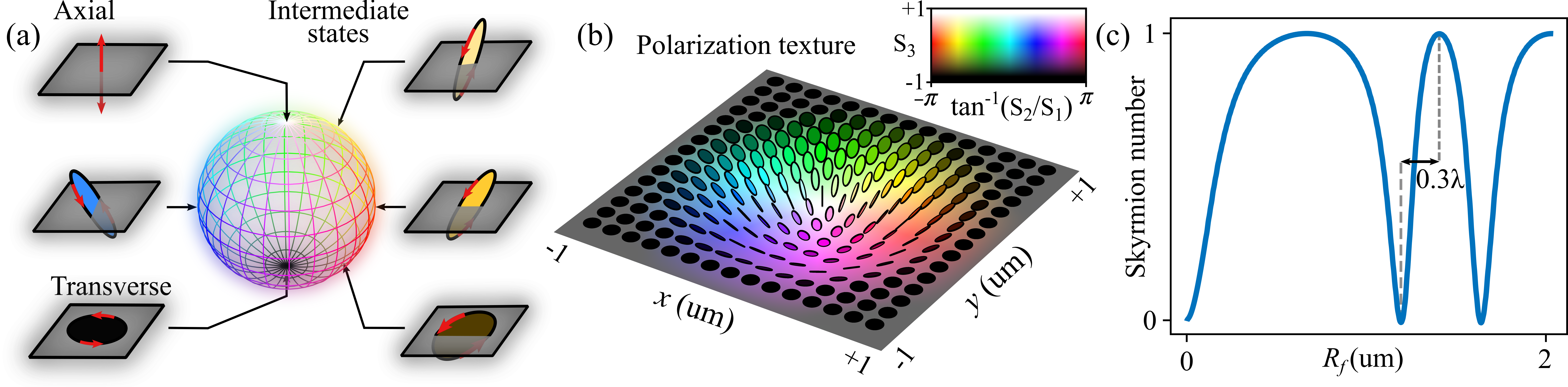}
    \caption{(a) the TA-PS representing all the states between transverse (RCP, south pole) and longitudinal polarization (axial, north pole). (b) Skyrmionic polarization texture at the focal plane of a focused EVB with $s=1$, $\mu=0.2$, $\sigma=\SI{25}{\micro\metre}$ of Fig. \ref{fig1}(c). The color map represents the coordinates $(S_1,S_2,S_3)$ in the TA-PS. Integration of (\ref{eq:Nsk_xy}) over the displayed region yields $N_{\rm sk}=1$. (c) Skyrmion number in Eq. (\ref{eq:Nsk_xy}) as a function of the truncation radius $R_f$ for the focused flat-top vortex with $s=1$ in Fig. \ref{fig1}(d). Data: $f=\SI{1500}{\micro\metre}$, $\lambda=\SI{735}{\nano\metre}$, $\rrm{NA}=0.5$.}
    \label{fig2}
\end{figure*}

Instead, GS skyrmions are naturally present in an OV without resorting to such superpositions. As seen in Fig. \ref{fig1}(b) for linearly polarized EVBs $\psi_x{\boldsymbol u}_x=\psi(r)e^{is\varphi}{\boldsymbol u}_x$ with $s=1$, the focused field develops a relevant longitudinal component at the phase singularity, where the transversal component vanishes. The transversal component maintains its central vortex, and the phase of the longitudinal component is nearly flat in its central intensity lobe, followed by a binary vortex constellation along the $x$ axis, where the axial field vanishes. Thus, in a circle of diameter equal to the distance between the first two vortices, all relative intensities $\lr{|\psi_{z}|^2-|\psi_x|^2}/\lr{|\psi_{z}|^2+|\psi_x|^2}$ are present, and all relative phases $\rrm{arg}\lr{\psi_{z}}-\rrm{arg}\lr{\psi_x}$ are also present by virtue of the azimuthal phase variation of the transversal vortex, yielding the polarization structure of a Stokes skyrmion of polarization in a sagittal plane in this region of the focal plane. Indeed, evaluation of the skyrmion number with the usual definition of the Stokes parameters but for the $z$ and $x$ components yields $N_{\rm SK} =1$ integrating in this circle.

\begin{figure}[b]
    \centering
    \includegraphics[width=\linewidth]{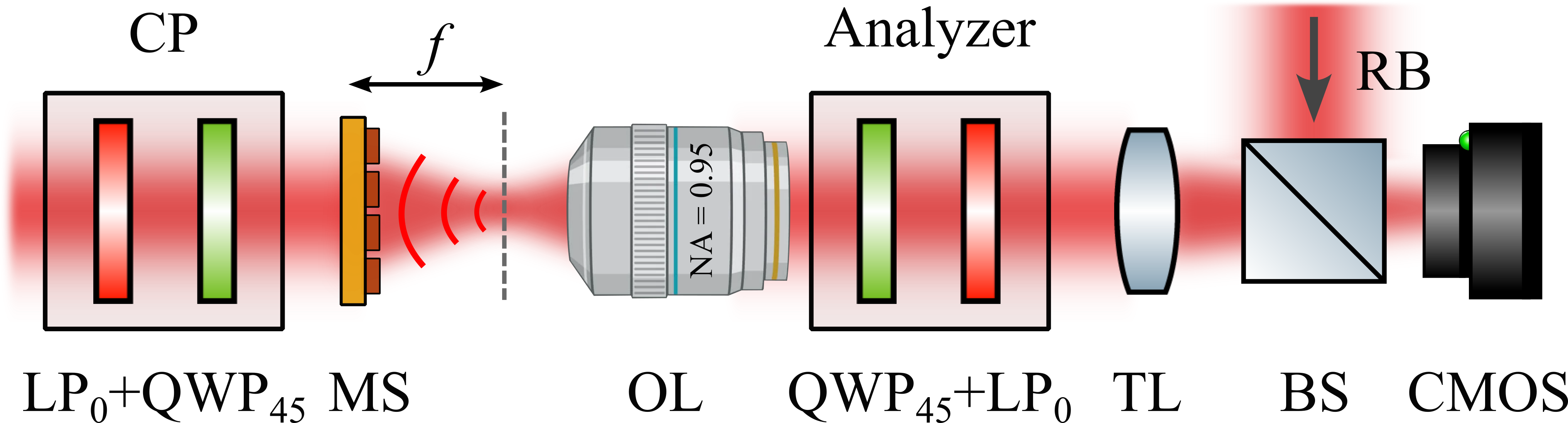}
    \caption{Schematic setup for complex amplitude reconstruction. 
    $\rrm{LP}_\alpha$: polarizer rotated by $\alpha$, $\rrm{QWP}_\alpha$: quarter-wave plate rotated by $\alpha$, OL: objective lens, TL: tube lens, BS: non-polarizing beam splitter, RB: reference beam.}
    \label{fig3}
\end{figure}

With CP,
$\psi_\perp {\boldsymbol u}_{l,r}=\psi(r)e^{is\varphi}{\boldsymbol u}_{l,r}$, the longitudinal field is symmetric and more intense both for focused EVB and uniform profiles [Figs. \ref{fig1}(c) and (d)], and does not carry any vorticity for $s=1$ and RCP, and for $s=-1$ and LCP. The phase profiles at the focal plane are those of a vortex beam for the transversal component and a flat wavefront for the longitudinal component. The $\pi-$phase jumps reveal the points where the longitudinal field vanishes, similar to a LG beam with nonzero radial index. In the region delimited by the first $\pi-$phase jump, the two components $\psi_{\perp}$ and $\psi_z$ cover again all the relative phases, 
courtesy of the TC difference in (\ref{eq:CP_profile}), and also all the relative intensities. 

In fact, these focal profiles form skyrmions of TA polarization, covering all states of TA polarization at the focal plane. We define the TA-Stokes parameters as
\begin{eqnarray}
S_0 &=& |\psi_\perp|^2+|\psi_z|^2,\quad
S_1 = 2{\rm Re}\{\psi_\perp^{\star}\psi_z\}, \nonumber \\
S_2 &=& -2{\rm Im}\{\psi_\perp^{\star}\psi_z\}, \quad
S_3 =|\psi_z|^2-|\psi_\perp|^2,
\end{eqnarray}
with $\psi_\perp$ the transverse circularly polarized component (RCP in this case) and $\psi_z$ the longitudinal component. The normalized vector field $\bm{s}=(S_1,S_2,S_3)/S_0$ defines the coordinates on the surface of the TA Poincaré sphere (TA-PS) depicted in Fig. \ref{fig2}(a), which represents all states of TA polarization.
At the south pole of this parametric sphere,
the electric field 
oscillates in the $x$-$y$ plane with RCP. For increasing latitude, the transverse circle transitions to an off-plane ellipse (intermediate TA polarization), and 
degenerates into a longitudinal line at the north pole (axial polarization). The phase difference between $\psi_\perp$ and $\psi_z$ determines the 
azimuth 
of the polarization plane 
[Eq. (S3) in SI], whereas their relative intensities 
defines the inclination angle of the polarization plane with respect to the $z$-axis [Eq. (S4) in SI]. We stress that here the transverse polarization is uniform (scalar beam), thus the full polarization is described by two degrees of freedom (2-sphere) due to the axial field.

As an example, we depict the polarization profile of the focused EVB ($\rrm{NA}=0.5$) in Fig. \ref{fig2}(b), showing the full coverage of the surface of the TA-PS at the focal plane. The lightness-hue colormap shows the corresponding coordinates at the TA-PS (see inset). This texture of the GS skyrmion is a TA counterpart of Stokes skyrmions.
As a validation of this topological polarization profile we obtain a skyrmion number of $N_{\rm SK}=1$ when integrating (\ref{eq:Nsk_xy}) up to the radius  $R_f\approx\SI{0.86}{\micro\metre}$, of the first $\pi-$phase jump in the axial component (south pole in the TA-PS). 

Similar features are found in 
the focused flat-top vortex. The intensity and phase profiles are presented in Fig. \ref{fig1}(d) for a numerical aperture $\rrm{NA}=0.5$. The transversal component experiences multiple $\pi-$phase jumps in addition to the $2\pi$ azimuthal variation, and the longitudinal component presents a flat phase except $\pi$-steps at its zeros. In Fig. \ref{fig2}(c) we show that in contrast with EVBs, the multiple zeros of the transverse field yield an intricate GS skyrmionium texture with a skyrmion number with deep sub-wavelength oscillation from 1 to 0.
\begin{figure*}[t!]
    \centering
\includegraphics[width=0.95\linewidth]{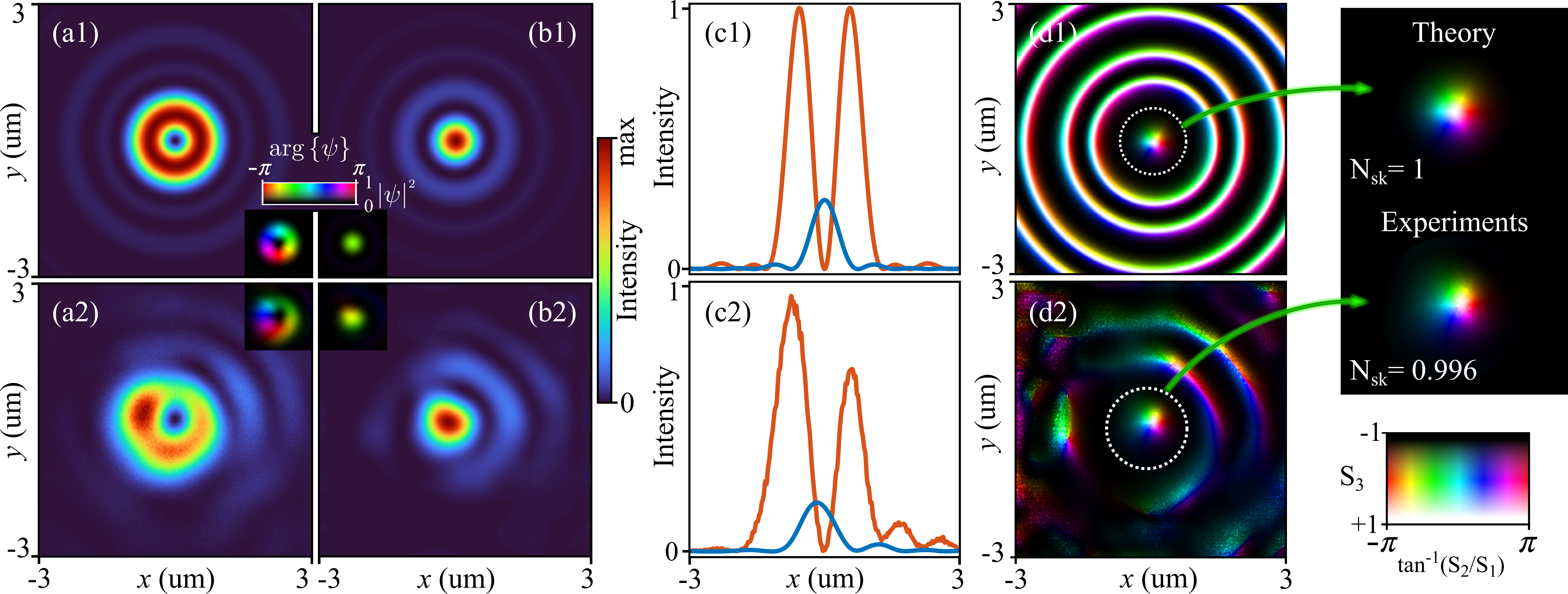}
    \caption{Experimental observation of TA skyrmionic beams: transverse (a2) and longitudinal (b2) intensity profiles at $z=f$, and their corresponding theoretical profiles (a1) and (b1). The inset shows the intensity-weighted phase. The intensity profiles for $y=0$ are presented in (c1) and (c2) for both theory and experiments. The theoretical (d1) and experimental (d2) TA Stokes textures and zooming textures of the central ring, yielding a skyrmion number $N_{\rm sk}=0.996$ in experiments. Data: output beam after the MS is a converging RCP-polarized uniform vortex with $s=1$, $f=\SI{433}{\micro\metre}$, $\rrm{NA}=0.5$, $D_{\rm MS}=\SI{500}{\micro\metre}$, $\lambda=\SI{735}{\nano\metre}$.}
    \label{fig4}
\end{figure*}

Observation of GS skyrmions 
requires an axial field with a significant peak intensity compared to that of the transverse field. Small transverse profiles are required to enhance the derivatives in (\ref{eq:CP_profile}). In the experiment, 
focusing the flat-top vortex with a moderate but paraxial $\rrm{NA}=0.5$ suffices.
For compactness, we carry out both wavefront vortex shaping and focusing with a single metasurface (MS). 
MSs are ultra-thin, 2D arrays of nanostructures that enable precise control over the properties of light~\cite{meta_gen_5,kuznetsov2024roadmap,vortex_2}.
The MS used here, with a diameter of $D_{\rm MS} = \SI{500}{\micro\metre}$, manipulates the phase based on the geometric phase, generating a vortex with $s=1$, wave front curvature $f=\SI{433}{\micro\metre}$, uniform intensity, and RCP. More details of the MS design are found in Sec. S2 of SI.

In the setup of Fig. \ref{fig3}, we prepare LCP impinging on the MS, which converts it into RCP, shapes the vortex and focuses it.
The focal plane is imaged using a microscope of $125\rrm{X}$ magnification and the cross-polarized wave (RCP) is selected with an analyzer, removing any residual co-polarized wave (LCP). Then, the signal beam interferes with the reference beam and the interference fringes are recorded in a CMOS detector~\cite{bone1986fringe}. This allows us to retrieve the full transverse complex field $\lr{E_x,E_y}$, which uniquely determines the longitudinal component through Gauss's law [Eq. (S5) of SI]~\cite{herrera2023measurement,measure_Ez}. 

The experimentally reconstructed transversal and axial electric field intensities and phases (insets) are presented in Fig. \ref{fig4}(a2)-(b2), which are in agreement with the theoretical prediction in (a1)-(b1). 
The intensity profiles at $y=0$ for the transverse (red) and longitudinal (blue) fields are depicted in (c1) and (c2) for theory and experiments, respectively. Finally, in (d1)-(d2) we present the theoretical and measured TA polarization profile, which yields an experimental skyrmion number of $N_{\rm exp}=0.996$ for the central ring, in great agreement with the theoretical value $N_{\rm th}=1$.

The generality of GS skyrmions in OVs is demonstrated by taking the solution to the wave equation for monochromatic light $\psi_\perp = r^{|s|}e^{is\varphi}$ that captures the essence of the vortex, wherever it is nested. Eq. (\ref{eq:CP_profile}) yields $\psi_z=i r^{|s|-1}a|s| e^{\pm i(|s|-1)\varphi}$ for the respective ($s>0$, RCP) and ($s<0$, LCP), where $a=\sqrt{2}/k$. It is straightforward to evaluate analytically the TA-Stokes parameters, the normalized vector field $\textbf{s}$, and its derivatives, to obtain a skyrmion density 
$n_{\rm SK} = \pm a^2s^2/[\pi(a^2s^2+r^2)^2],$
that integrates in the whole transversal plane to $N_{\rm SK} = \pm 1$, regardless of the magnitude of the TC. Of course, this is an idealized situation. When the OV is embedded in a beam, the GS skyrmion becomes localized. 

In conclusion, 
we have unveiled the existence of a new type of optical Stokes skyrmion surrounding the phase singularity of optical vortices when accounting for the unavoidable axial field component derived from Gauss's divergence law. We have shown that they can emerge up to observability with standard characterization techniques even with paraxially focused fields with an enhanced axial component.
Gauss-Stokes skyrmions present pure longitudinal polarization at the vortex singularity, which is encircled by a continuous transverse-axial polarization pattern fulfilling a topological map from the transverse-axial Poincaré sphere to the focal plane. We have indeed observed such transverse-axial skyrmionic textures using a geometric phase metasurface and verified the predicted topological features. 
Our findings shed light on the intimate nature of the phase singularity of optical vortices, placing it in intrinsic relationship with the emerging family of structured light known as optical skyrmions.

\textbf{Acknowledgments.}
The authors acknowledge no conflict of interest. N.M.-C. thanks Xie Xi \href{https://orcid.org/0000-0003-2748-7974}{\orcidicon{8pt}} for providing helpful plotting tools.

\textbf{Author contributions.}
N.M.-C and M.A.P. conceived the idea and carried out the numerical simulations. D.K.S. performed the fabrication of the MS. D.K.S. conducted the experimental part with N.M.-C., who contributed to the characterization, and R.P.-D., who contributed to the design of the experiment. N.M.-C. and M.A.P. co-wrote the first draft of the paper with inputs from Y.S. All authors participated in the analysis of the results and discussions.

\textbf{Funding.}
N.M.-C. acknowledges support from Nanyang Technological University under SINGA Scholarship. Y.S. acknowledges support from Nanyang Technological University Start Up Grant, Singapore Ministry of Education (MOE) AcRF Tier 1 grant (RG157/23), MoE AcRF Tier 1 Thematic grant (RT11/23), and Imperial-Nanyang Technological University Collaboration Fund (INCF-2024-007).  This work was supported in part by the AME Programmatic Grant, Singapore, under Grant A18A7b0058. M.A.P. acknowledges support from the Spanish Ministry of Science and Innovation, Gobierno de Espa\~na, under Contract No. PID2021-122711NB-C21.

\bibliography{bibliography}

\end{document}


\title{Supplementary Information for Optical Skyrmions of Vortex Darkness}

\maketitle
\section{Geometrical description of the polarization features}
\begin{figure}
    \centering
\includegraphics[width=0.5\linewidth]{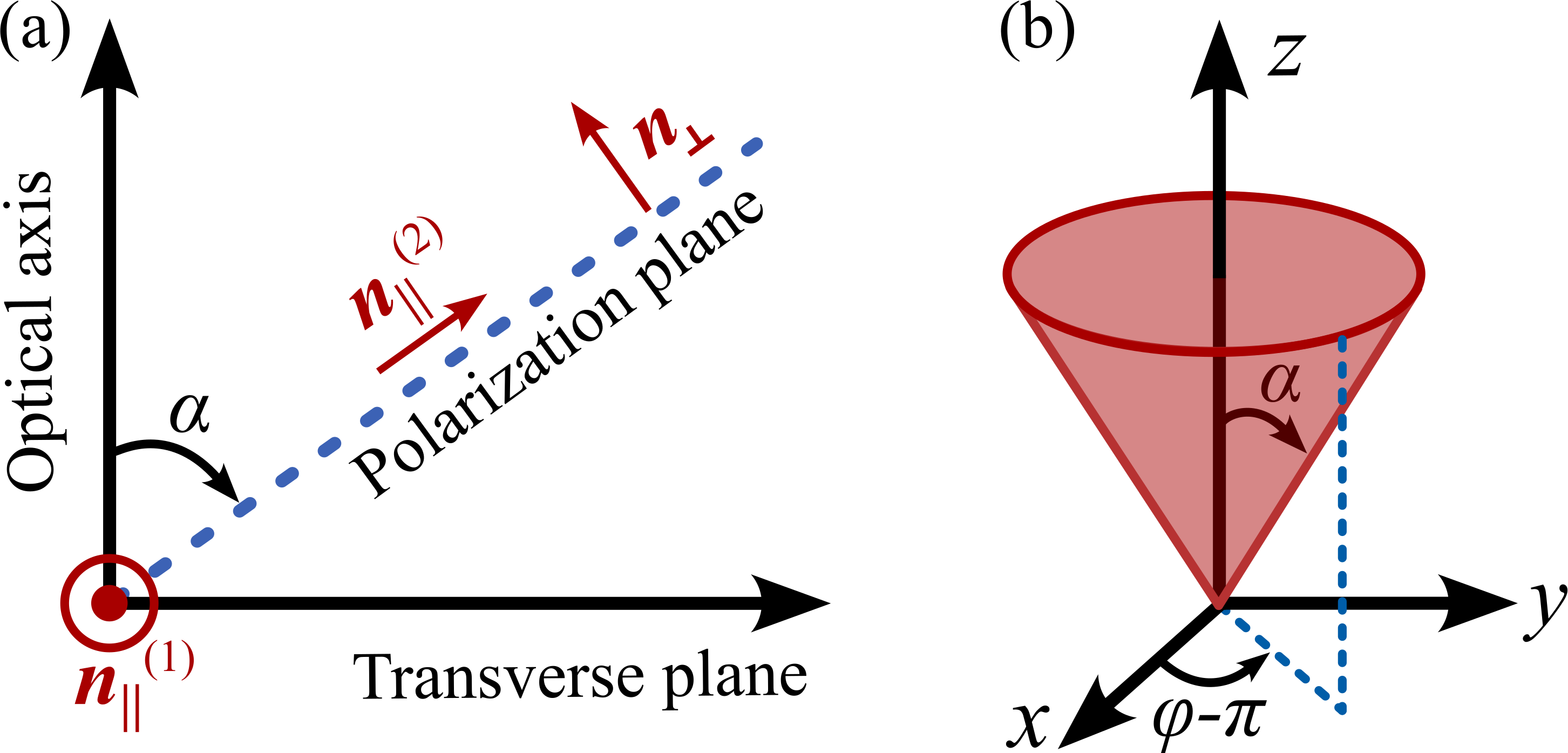}
    \caption{(a) conical angle $\alpha$ of the polarization plane with respect to the optical axis $z$ and the auxiliary unit vectors along the plane and normal to the plane. Azimuthal angle $\varphi$ of the polarization plane for a fixed inclination angle $\alpha$ given a fixed axial-to-transverse electric field amplitudes.}
    \label{figS1}
\end{figure}

We can describe the temporal evolution of the local electric field as 
\begin{equation}
\bm{E}=
\begin{pmatrix}
E_x \\ E_y  \\ E_z
\end{pmatrix}
=\begin{pmatrix}
E_\perp\cos{(\omega t)} \\ E_\perp\sin{(\omega t)}  \\ E_\parallel\cos{(\omega t+\delta)}
\end{pmatrix},    
\end{equation}
where we have imposed a circularly polarized transverse field, $E_\perp$ and $E_\parallel$ are the transverse and longitudinal amplitudes and $\delta$ denotes their relative phase difference. In this case, we can express $\cos{(\omega t)}=E_x/E_0$ and $\sin{(\omega t)}=E_y/E_0$. Using basic trigonometric relations the longitudinal field can be expressed in terms of the transverse components as 
$$\frac{E_z}{E_\parallel}=\frac{E_x}{E_\perp}\cos{\delta}-\frac{E_y}{E_\perp}\sin{\delta},$$
which defines the equation of the polarization plane. The unit vector normal to this plane is simply
\begin{equation}\label{eq:n_t}
    \bm{n}_\perp=\frac{\ez}{\sqrt{\et^2+\ez^2}}\lr{-\cos\delta,\sin\delta,\frac{\et}{\ez}},
\end{equation}
which yields $\bm{n}_\perp=\lr{0,0,1}$ for $\ez=0$ (transverse polarization), and an undetermined transverse vector $\bm{n}_\perp=\lr{-\cos{\delta},\sin{\delta},0}$ for $\et=0$ (longitudinal polarization), since relative phase $\delta$ is undefined. We can evaluate the orientation of the polarization plane with respect to the $XY$ plane by projecting the vector (\ref{eq:n_t}). With this we obtain the \textit{azimuth} of the polarization plane, given by
\begin{equation}\label{eq:azimuth}
    \varphi=\tan^{-1}{\lr{-\frac{\sin{\delta}}{\cos{\delta}}}}=-\delta,
\end{equation}
see Fig. \ref{figS1}(b). This azimuth becomes undefined at the vortex singularity, yielding a point of pure longitudinal polarization (Fig. 2(b) in main text). In order to determine the elevation of the polarization plane with respect to the transverse plane \ref{figS1}(a) we follow a simple geometric approach. First, we intersect the polarization plane with the transverse plane, yielding the following line 
$$\frac{E_x}{E_\perp}\cos{\delta}-\frac{E_y}{E_\perp}\sin{\delta}=0.$$
The unit vector along such a line can be evaluated straightforwardly as
$$\bm{n}_\parallel^{(1)}=\lr{-\sin{\delta},-\cos{\delta},0}.$$
With this unit vector and $\bm{n}_\perp$ we can obtain another unit vector $\bm{n}_\parallel^{(2)}$ parallel to the polarization plane and orthogonal to the previous two vectors (see \ref{figS1}(a)) through the simple cross product
$$\bm{n}_\parallel^{(2)}=\bm{n}_{\perp}\times\bm{n}_\parallel^{(1)}=\frac{\et}{\sqrt{\et^2+\ez^2}}\lr{\cos{\delta},-\sin{\delta},\frac{\ez}{\et}}.$$
Finally, we evaluate the \textit{inclination angle} $\alpha$ of the polarization plane with respect to the $z-$axis through the dot product $\bm{n}_\parallel^{(2)}\cdot\bm{E}_z$ yielding
\begin{equation}\label{eq:tilt}
    \alpha=\cos^{-1}{\lr{\bm{n}_\parallel^{(2)}\cdot\bm{E}_z}}=\cos^{-1}{\lr{\frac{\ez}{\sqrt{\et^2+\ez^2}}}}.
\end{equation}
Eq. (\ref{eq:tilt}) yields $\alpha=\pi/2$ for $\ez=0$ (transverse polarization) and $\alpha=0$ for $\et=0$ (longitudinal polarization). 
\section{Metasurface design and fabrication}
\begin{figure*}[b!]
    \centering
    \includegraphics[width=0.9\linewidth]{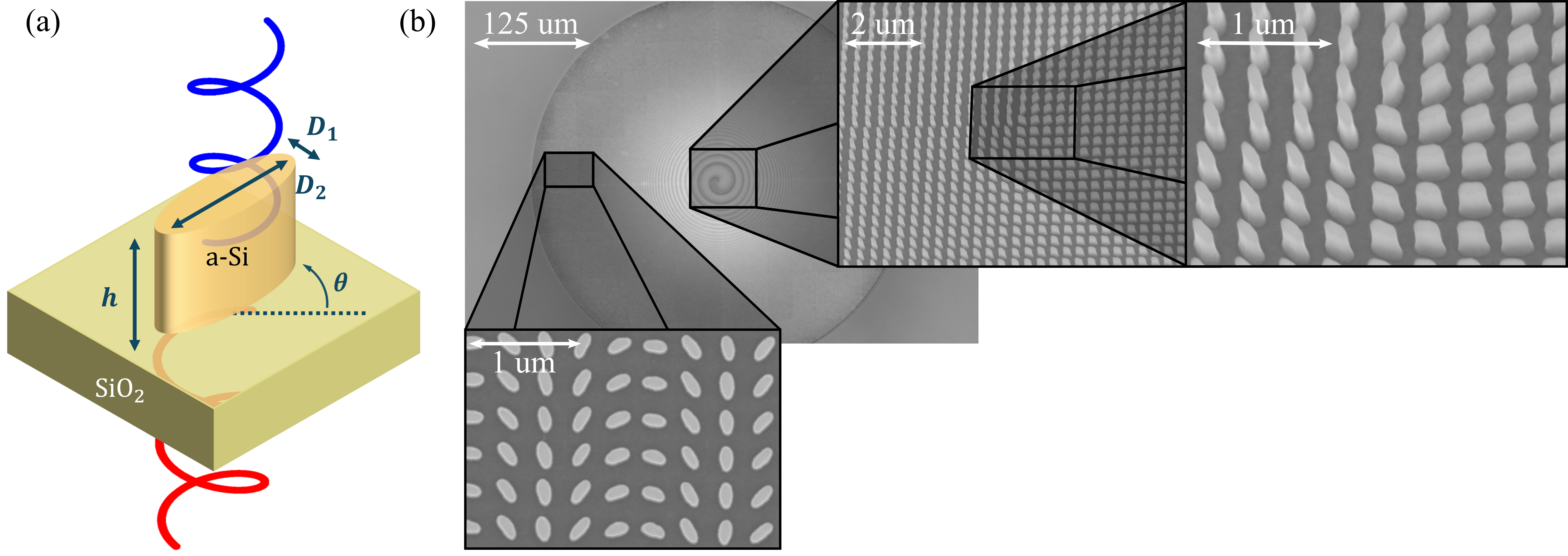}
\caption{(a) MS unit cell (period $P=\SI{325}{\nano\metre}$) constituted by an amorphous-Silicon (a-Si) elliptical meta-atom laying on top on an $\SI{23}{\nano\metre}$-thick ITO coated glass substrate, with constant geometry $h=\SI{350}{\nano\metre}$, $D_1=\SI{90}{\nano\metre}$, $D_2=\SI{250}{\nano\metre}$ and varying rotation angle $\theta$. Optical image of the MS (b, top-left) and SEM images showing meta-atom rotation inducing optical vortex and beam fousing: top views (bottom) and 30$\degree$-tilted views (top-center and top-left).}
    \label{figS2}
\end{figure*}

We design a standard geometric phase MS using commercial software Lumerical FDTD. The unit cell is square with a period $P=\SI{325}{\nano\metre}$ in which an amorphous-Silicon elliptic meta-atom lays on top of a quartz substrate. The meta-atom geometry is fixed with minor and major axes $D_1=\SI{90}{\nano\metre}$ and $D_2=\SI{250}{\nano\metre}$ respectively, which have been chosen to achieve maximum cross-polarization conversion for input circularly polarized light with wavelength $\lambda=\SI{735}{\nano\metre}$. The output light is cross-polarized and its phase is uniquely manipulated by the rotation angle $\rrm{exp}\lr{i2\theta}$. The encoded phase profile $\rrm{exp}\lr{i\Phi(\bm{r})}$ can be expressed as
$$\Phi(\bm{r})=s\varphi-k\sqrt{f^2+x^2+y^2},$$
with $f=\SI{433}{\micro\metre}$ the focal length chosen for achieving $\rrm{NA}=0.5$ for a MS diameter of $D_{\rm MS}=\SI{500}{\micro\metre}$, and $s=1$ the topological charge of the vortex. We stress that no amplitude modulation is included in the MS in order to generate a flat-top vortex with converging wavefront. A schematic picture of the unit cell is shown in Fig. \ref{figS2}(a), where the input LCP light is converted into RCP light in transmission. An optical image of the MS is shown in Fig. \ref{figS2}(b), together with Scanning Electron Microscope (SEM) images of the center and edges of the MS for different magnifications. MS is fabricated on a $\SI{23}{\nano\metre}$-thick ITO coated glass substrate where amorphous-Silicon (a-Si) of $\SI{350}{\nano\metre}$ thickness is deposited using Inductively Coupled Plasma Chemical Vapour Deposition (ICPCVD) method. In the following steps, MS is patterned using single step electron-beam lithography and dry etched in an inductively coupled plasma system.

\section{Experimental retrieval of the axial component}
As detailed in previous works~\cite{measure_Ez,herrera2023measurement}, measurement of the complex transverse fields $(E_x,E_y)$ uniquely defines the axial field through Gauss's divergence $\nabla\cdot\bm{E}=0$. In our analysis we have used the strict version of Gauss's law and not the paraxial approximation to probe that the described phenomena are not only supported by the paraxial theory. This equation can be easily solved in the spatial frequency domain with the solution in real space given by
\begin{equation}
    E_z(x,y)=-\mathcal{F}^{-1}\lr{\frac{k_x\hat{E}_x(k_x,k_y)+k_y\hat{E}_y(k_x,k_y)}{\sqrt{k^2-k_x^2-k_y^2}}},
\end{equation}
where $\mathcal{F}^{-1}$ denotes the inverse 2D Fourier transform, $\hat{E}_x$ and $\hat{E}_y$ are the Fourier transforms of the $x$- and $y$-polarized electric fields obtained by interferometry, and $k=\omega/c$ is the wavenumber.

\bibliography{bibliography}